\documentclass[prb,a4paper,12pt,onecolumn,superscriptaddress,amsmath,amsfonts,amssymb,preprintnumbers,citeautoscript]{revtex4}
\pdfoutput=1
\usepackage[T1]{fontenc}\usepackage[latin1]{inputenc}
\usepackage{graphicx,color,booktabs,microtype,afterpage,sidecap,ulem}
\usepackage[charter]{mathdesign}

\usepackage[colorlinks,plainpages=false,linkcolor=black,urlcolor=blue,citecolor=black,pdfpagemode=UseNone,pdfstartview=FitBH]{hyperref}

\makeatletter
\renewcommand{\@evenfoot}{\hfill\raisebox{-3em}{\bf\thepage}\hfill}
\renewcommand{\@oddfoot}{\hfill\raisebox{-3em}{\bf\thepage}\hfill}
\makeatother

\begin{document}

\title{Intense paramagnon excitations \\ in a large family of high-temperature superconductors}

\author{M. Le Tacon}
%\email[email: ]{m.letacon@fkf.mpg.de}
\affiliation{Max Planck Institute for Solid State Research, 70569 Stuttgart, Germany}

\author{G. Ghiringhelli}
\affiliation{CNR-SPIN, Dipartimento di Fisica, Politecnico di Milano, I-20133 Milano, Italy}

\author{J. Chaloupka}
\affiliation{Max Planck Institute for Solid State Research, 70569 Stuttgart, Germany}

\author{M. Moretti Sala}
\affiliation{CNR-SPIN, Dipartimento di Fisica, Politecnico di Milano, I-20133 Milano, Italy}

\author{V. Hinkov}
\affiliation{Max Planck Institute for Solid State Research, 70569 Stuttgart, Germany}
\affiliation{Department of Physics and Astronomy, University of British Columbia, Vancouver, Canada V6T1Z1}

\author{M. W. Haverkort}
\affiliation{Max Planck Institute for Solid State Research, 70569 Stuttgart, Germany}

\author{M. Minola}
\affiliation{CNR-SPIN, Dipartimento di Fisica, Politecnico di Milano, I-20133 Milano, Italy}

\author{M. Bakr}
\affiliation{Max Planck Institute for Solid State Research, 70569 Stuttgart, Germany}

\author{K. J. Zhou}
\affiliation{Swiss Light Source, Paul Scherrer Institut, CH-5232 Villigen PSI, Switzerland}

\author{S. Blanco-Canosa}
\affiliation{Max Planck Institute for Solid State Research, 70569 Stuttgart, Germany}

\author{C. Monney}
\affiliation{Swiss Light Source, Paul Scherrer Institut, CH-5232 Villigen PSI, Switzerland}

\author{Y. T. Song}
\affiliation{Max Planck Institute for Solid State Research, 70569 Stuttgart, Germany}

\author{G. L. Sun}
\affiliation{Max Planck Institute for Solid State Research, 70569 Stuttgart, Germany}

\author{C. T. Lin}
\affiliation{Max Planck Institute for Solid State Research, 70569 Stuttgart, Germany}

\author{G. M. De Luca}
\affiliation{CNR-SPIN, Complesso Monte Santangelo via Cinthia, I-80126 Napoli, Italy}

\author{M. Salluzzo}
\affiliation{CNR-SPIN, Complesso Monte Santangelo via Cinthia, I-80126 Napoli, Italy}

\author{G. Khaliullin}
\affiliation{Max Planck Institute for Solid State Research, 70569 Stuttgart, Germany}

\author{T. Schmitt}
\affiliation{Swiss Light Source, Paul Scherrer Institut, CH-5232 Villigen PSI, Switzerland}

\author{L. Braicovich}
\affiliation{CNR-SPIN, Dipartimento di Fisica, Politecnico di Milano, I-20133 Milano, Italy}

\author{B. Keimer}
%\email[email: ]{b.keimer@fkf.mpg.de}
\affiliation{Max Planck Institute for Solid State Research, 70569 Stuttgart, Germany}

\begin{abstract}
\center\bigskip\thispagestyle{plain}
\begin{minipage}{\textwidth}\textbf{In the search for the mechanism of high-temperature superconductivity, intense research has been focused on the evolution of the spin excitation spectrum upon doping from the antiferromagnetic insulating to the superconducting states of the cuprates.\cite{Abanov03,Birgeneau06,Eschrig06} Because of technical limitations, the experimental investigation of doped cuprates has been largely focused on low-energy excitations in a small range of momentum space.\cite{Fong96, Hayden_Nature2004, Hinkov_Nature04, He_Science02, Fauque_PRL2006, Reznik_PRB2008, Xu_NaturePhysics2009, Yu_PRB2010} Here we use resonant inelastic x-ray scattering~\cite{Braicovich_PRL2009,Schlappa_PRL2009, Guarise, Braicovich_PRB2010} to show that a large family of superconductors, encompassing underdoped YBa$_2$Cu$_4$O$_8$ and overdoped YBa$_2$Cu$_3$O$_{7}$, exhibits damped spin excitations (paramagnons) with dispersions and spectral weights closely similar to those of magnons
  in undoped cuprates.
%The results are in excellent agreement with the spin excitations obtained by exact diagonalization of the $\bf t-J$ Hamiltonian on finite-sized clusters.
The comprehensive experimental description of this surprisingly simple spectrum permits quantitative tests of magnetic Cooper pairing models. A numerical solution of the Eliashberg equations for the magnetic spectrum of YBa$_2$Cu$_3$O$_{7}$ reproduces its superconducting transition temperature within a factor of two, a level of agreement comparable to Eliashberg theories of conventional superconductors.~\cite{Carbotte}
}\end{minipage}
\end{abstract}

\maketitle\thispagestyle{empty}\clearpage

The twenty-fifth anniversary of the discovery of high-temperature superconductivity is approaching without a clear and compelling theory of the mechanism underlying this phenomenon. After the discovery of an unconventional ($d$-wave) symmetry of the Cooper pair wave function in the copper oxides, the thrust of research has been focused on the role of repulsive Coulomb interactions between conduction electrons, which naturally explain this pairing symmetry. However, since even simple models based on repulsive interactions %(such as the Hubbard and $t-J$ models)
have thus far defied a full solution, the question of whether such interactions alone can generate high-temperature superconductivity is still open. A complementary, more empirical approach has asked whether antiferromagnetic spin fluctuations, which are a generic consequence of Coulomb interactions, can mediate Cooper pairing in analogy to the phonon-mediated pairing mechanism in conventional superconductors~\cite{Abanov03}.
This scenario requires the existence of well-defined antiferromagnetic spin fluctuations in the superconducting range of the cuprate phase diagram (for mobile hole concentrations $5\% \leq p \leq 25\%$ per copper atom), well outside the narrow stability range of antiferromagnetic long-range order ($0 \leq p \leq 2\%$).

An extensive series of experiments using inelastic spin-flip scattering of neutrons has indeed revealed low-energy spin fluctuations in doped
cuprates~\cite{Birgeneau06}. Signatures of coupling between spin and charge excitations have also been identified~\cite{Eschrig06}, and evidence has been reported that this coupling is strong enough to mediate superconductivity in underdoped cuprates.~\cite{Dahm_NaturePhysics2009}
However, for cuprate compounds hosting the most robust superconducting states, namely those that are optimally doped to exhibit $T_c \geq 90$ K, inelastic neutron scattering (INS) experiments have thus far mostly revealed spin excitations over a narrow range of excitation energies $E \sim 30-70$ meV, wave vectors ${\bf Q}$ covering only $\sim 10$ \% of the Brillouin zone area around the antiferromagnetic ordering wave vector ${\bf Q}_{AF}$, and temperatures $T < T_c$~\cite{Fong96, Hayden_Nature2004, Hinkov_Nature04, He_Science02, Fauque_PRL2006, Reznik_PRB2008, Xu_NaturePhysics2009, Yu_PRB2010}.
The energy- and momentum-integrated intensity of these excitations constitutes only a few percent of the spectral weight of spin waves in antiferromagnetically ordered cuprates~\cite{Eschrig06, Kee02}, and is thus clearly insufficient to support high-$\rm T_c$ superconductivity.
Although the recent discovery of a weakly dispersive magnetic excitation in the model system HgBa$_2$CuO$_{4+\delta}$ with $E \sim 50$ meV may account for some of the missing spectral weight~\cite{Li_Nature2010}, the apparent weakness of antiferromagnetic fluctuations in optimally doped compounds has been used as a central argument against magnetically mediated pairing scenarios for the cuprates \cite{Maksimov10}.

This picture is, however, strongly influenced by technical limitations of the INS method that arise from the small cross section of magnetic neutron scattering in combination with the weak primary flux of currently available high-energy neutron beams. Because of intensity constraints, even the detection of undamped spin waves in antiferromagnetically ordered cuprates over their full band width of $\sim 300$ meV has required single-crystal samples with volumes of order 10 cm$^3$, which are very difficult to obtain~\cite{Tranquada_PRB1989,Reznik_PRB1996,Hayden_PRB1996,Coldea_PRL01}. Doping further reduces the intensity of the INS profiles and exacerbates these difficulties.

Here we take advantage of recent progress in the development of high-resolution resonant inelastic x-ray scattering (RIXS) in order to explore magnetic excitations in a wide energy-momentum window that has been largely hidden from view by INS. Experiments on undoped cuprates have shown that RIXS with photon energies at the Cu $L_3$ absorption edge is sensitive to single-magnon excitations~\cite{Braicovich_PRL2009,Schlappa_PRL2009, Guarise, Braicovich_PRB2010} by virtue of the strong spin-orbit coupling of the $2p_{3/2}$ core-hole intermediate state~\cite{Ament_PRL2009, Haverkort}, in excellent agreement with INS measurements~\cite{Coldea_PRL01, Braicovich_PRL2010}. Initial RIXS experiments on doped La$_{2-x}$Sr$_{x}$CuO$_4$ have revealed a complex two-component lineshape with excitations extending up to about 350 meV; the presence of two branches was attributed to phase separation between the superconducting state and a competing state with incommensurate spin and charge orde
 r~\cite{Braicovich_PRL2010}. We have applied the same method to YBa$_2$Cu$_3$O$_{6+x}$ (YBCO$_{6+x}$), Nd$_{1.2}$Ba$_{1.8}$Cu$_3$O$_{6+x}$ (NdBCO$_{6+x}$), and YBa$_2$Cu$_4$O$_8$, which are much less affected by doping-induced disorder and do not show phase separation~\cite{Bobroff_PRL02}.

Although one cannot reach $Q_{AF}$ with RIXS, one of its great advantages over INS is that it allows measurements of magnetic excitations with sizable intensity over much of the accessible reciprocal space, even on very small sample volumes. The results presented here have been obtained on thin films and on millimeter-sized single crystals far below the volume requirements of INS (see Supplementary Information). Figure 1 shows a sketch of the scattering geometry of our experiment, as well as RIXS spectra obtained on undoped antiferromagnetic NdBCO$_6$ (Fig. 1 c,d) and underdoped superconducting NdBCO$_7$ (Fig. 1 f,g),
%as well as  underdoped YBa$_2$Cu$_3$O$_{6.6}$ (YBCO$_{6.6}$, $T_c$ = 61 K)
for incident photon polarizations in and out of the scattering plane ($\pi$ and $\sigma$ geometries) and a momentum transfer $q_{//}$ along the reciprocal-space $a^*$ direction corresponding to 0.37 reciprocal lattice units (r.l.u.). For both scattering geometries, the spectra exhibit an intense peak located around 1.7 eV energy loss that arises from optically forbidden \emph{dd} excitations (that is, transitions of the unpaired hole of Cu$^{2+}$ from the $d_{x^2-y^2}$ to other $d$ orbitals)~\cite{Ghiringhelli_PRL2004} on top of a continuum of charge-transfer excitations. At lower energies, we can see in both compounds an inelastic feature centered around 250 meV and a resolution-limited elastic peak.

For undoped antiferromagnetic NdBCO$_6$, we can decompose the response in the mid-infrared (MIR) region of the spectra (Fig. 1e) for energy losses below $\sim$ 500 meV following the method employed in Ref. \, \onlinecite{Braicovich_PRL2010}. In the $\pi$ scattering geometry, this leads to (i) an intense resolution-limited peak around 250 meV, (ii) a high-energy tail of this peak centered around 400 meV,  and (iii) a weak low-energy contribution around 100 meV. In the $\sigma$ scattering geometry, feature (i) is strongly suppressed. This polarization dependence allows us to assign this feature to a single-magnon excitation, in agreement with theoretical considerations~\cite{Braicovich_PRB2010, Ament_PRL2009} and previous investigations on other cuprates~\cite{Schlappa_PRL2009, Guarise, Braicovich_PRB2010,  Braicovich_PRL2010}. This assignment is further confirmed by the disappearance of the MIR inelastic response as the incident photon energy is moved away from the Cu-$L_3$ ab
 sorption edge (not shown). The weak features (ii) and (iii) are associated with higher-order spin excitations and lattice vibrations (single and multiple phonon excitations that are not individually resolved), respectively.

The energy of the single-magnon feature in NdBCO$_6$ depends strongly on $q_{//}$ (left panel of Fig. 2). Since the lineshape of the MIR spectrum is $q_{//}$-independent, the fitting procedure described above can be used to obtain the magnon dispersion relation. The result displayed in Fig. 3a is clearly different from the sinusoidal magnon dispersion observed in La$_2$CuO$_4$~\cite{Coldea_PRL01, Braicovich_PRL2010}.
Indeed, due to the presence of two antiferromagnetically coupled CuO$_2$ layers per unit cell, two magnon branches with different spin precession patterns are expected~\cite{Tranquada_PRB1989, Reznik_PRB1996, Hayden_PRB1996} for NdBCO$_6$: an acoustic branch with intensity $\propto \sin^2(\frac{d q_{\perp}}{2})$, where $q_{\perp}$ is the momentum transfer perpendicular to the CuO$_2$ layers and $d$ is the spacing between the two CuO$_2$ layers, and an optical branch with intensity $\propto \cos^2(\frac{d q_{\perp}}{2})$.
The result of the calculation of the relative intensities of these two branches for our scattering geometry is displayed in the inset of Fig.~3a. It shows that close to the Brillouin zone center the gapped optical branch dominates. Fitting our data using the spin-wave dispersion calculated for a bilayer in the framework of a simple Heisenberg model~\cite{Tranquada_PRB1989, Reznik_PRB1996, Hayden_PRB1996}, we obtain $J_{//}$=133 $\pm$ 2 meV and $J_{\perp}$=12 $\pm$ 3 meV for the intra- and inter-layer exchange constants, respectively, in excellent agreement with the values obtained in antiferromagnetic YBCO$_{6+x}$~\cite{Reznik_PRB1996,Hayden_PRB1996}.

%YBa$_2$Cu$_3$O$_{6.2}$ ($J_{//}$=120 meV and $J_{\perp}$= 13 meV~\cite{Reznik_PRB1996}) and in YBa$_2$Cu$_3$O$_{6.15}$ ($J_{//}$=125 $\pm$ 5 meV and $J_{\perp}$= 11 $\pm$ 2 meV~\cite{Hayden_PRB1996}).

We now turn to the doped systems. Figure~2 provides a synopsis of the experimental spectra for all systems investigated here, which span a wide range of doping levels: undoped NdBCO$_6$, strongly underdoped NdBCO$_7$ and YBCO$_{6.6}$ ($T_c$ = 65 K and 61 K, respectively), weakly underdoped YBa$_2$Cu$_4$O$_{8}$ ($T_c$ = 80 K), and weakly overdoped YBCO$_7$ ($T_c$ = 90 K). Because of their stoichiometric composition and electronic homogeneity, the latter two compounds have served as model compounds in the experimental literature on high-$T_c$ superconductivity, but apart from the ``resonant mode'' that appears in YBCO$_7$ below $T_c$ \cite{Fong96} no information has been available on their magnetic excitation spectra. Fig. 2 shows that broad MIR features are present at all doping levels in the same energy range as the single-magnon peak in undoped NdBCO$_6$. Since these features obey the same polarization dependence as the magnon mode (Figs. 1 f,h), they can be assigned to magn
 etic excitations.

Based on the metallic nature of the doped cuprates, one expects strong damping of magnetic excitations in the Stoner continuum of incoherent electron-hole excitations. We have therefore fitted these spectra to Voigt profiles that are the result of the convolution  of the Lorentzian lineshape of excitations with finite lifetime with the Gaussian resolution function. Since the fits yield excellent agreement with this simple profile (solid lines in Figs. 1 and 2), we can accurately extract the energies and half-widths-at-half-maximum (HWHM) of the magnetic excitation as a function of the transferred momentum (Fig.~3b,c). Before discussing these results, it is interesting to compare them to those previously reported on underdoped La$_{2-x}$Sr$_{x}$CuO$_4$~\cite{Braicovich_PRL2010}. As no double-peak structure is observed here, our data confirm the absence of phase separation in the YBCO and NdBCO families of compounds \cite{Bobroff_PRL02}.

The magnetic excitation energies of NdBCO$_7$, YBCO$_{6.6}$, and YBCO$_7$ at the Brillouin zone boundary are nearly identical
to the ones found in NdBCO$_6$. This implies that the in-plane exchange constant $J_{//}$ is not as strongly renormalized with doping as previously suggested based on extrapolation of lower-energy INS data~\cite{Lipscombe_PRL2009,Stock_PRB2005, Vignolle_07},  in agreement with the most recent INS findings on YBCO$_{6.5}$ using high energy neutrons~\cite{Stock_PRB2010}.
Upon approaching the $\Gamma$ point ($q_{//}=0$), we observe that the dispersion in the doped compounds is steeper than in NdBCO$_6$. Despite the fact that $\Gamma$ and ${\bf Q}_{AF}$ are no longer equivalent in the absence of magnetic long-range order, the RIXS data obtained on YBCO$_{6.6}$ nicely extrapolate to the low-energy ``hour glass'' dispersion around ${\bf Q}_{AF}$ previously extracted from INS data on samples prepared in an identical manner~\cite{Hinkov_natPhys2007}.
Since the slope of the upward-dispersing branch close to the magnetic zone center is flatter than the one of the antiferromagnetic spin-wave dispersion~\cite{Stock_PRB2005, Hinkov_natPhys2007}, this indicates the presence of an inflexion point in the dispersion of magnetic excitations in the doped compounds.
We note that the situation seems somewhat different for YBa$_2$Cu$_4$O$_{8}$, where the energy of the magnetic excitations close to the zone boundary is significantly reduced compared to the other systems ($\sim$ 210 meV instead of $\sim$ 300 meV), and the dispersion is flatter than in NdBCO$_6$. This may reflect different values of $J_{//}$ and $J_{\perp}$ in this system, and deserves further investigation.
The intrinsic HWHM of $\sim$ 200 meV of the inelastic signal extracted from our data (Fig.~3c) is much larger than the instrumental resolution and comparable to the magnon energies, indicating strong damping by Stoner excitations. The damping rate does not change substantially with $q_{//}$ and with doping.  Finally, while it is not possible yet to obtain absolute magnetic intensities from RIXS data, we can extract the relative intensity of the RIXS profiles by integrating the inelastic signal in the MIR region (see Supplementary Information). %, after self-absorption correction (see Supporting Material). The high-energy $dd$ excitation peak (Fig. 1 b,c) provide an additional calibration standard.
Remarkably, the integrated intensity obtained in this way is conserved upon doping from the antiferromagnetic insulator to the slightly overdoped superconductor.

%As shown in Fig.~3-C, the HWHM of the inelastic signal extracted from our data is much larger than the resolution of the experiment: after deconvolution, we obtain $\sim$ 200 meV throughout the whole BZ, a value found to be rather doping independent.It is also possible to extract from our data information about the evolution with doping of the intensity of the magnetic signal. It is not possible yet to get absolute magnetic intensities from RIXS data. However, we can extract the relative magnetic intensity by simply integrating the inelastic signal in the MIR region~\cite{absorption}. The results plotted in Fig.~3-D reveals that despite the obvious changes of lineshape, the overall magnetic intensity is \emph{conserved upon doping}, while going from AF-insulator to slightly overdoped superconducting cuprate.
We have thus demonstrated the existence of paramagnons, \emph{i.e.} damped but well-defined, dispersive magnetic excitations, deep in the Stoner continuum of cuprates with doping levels beyond optimal doping. Their spectral weights are similar to those of spin waves in the undoped, antiferromagnetically ordered parent material.

In order to obtain insight into the origin of this surprising observation, we have performed exact-diagonalization calculations of the $t-J$ Hamiltonian with $J/t = 0.3$ on finite-sized clusters, following the method proposed in Refs. \, \onlinecite{Tohyama_PRL1995, Eder_PRL1995, Dagotto_RMP1994}.
We used clusters with 18 and 20 spins arranged in various shapes to map the (H,K,0) reciprocal plane with sufficient momentum resolution (see Supplementary Information). The resulting spectra were convoluted by a Gaussian function with HWHM $0.1t$ in order to simulate the experimental resolution function. The calculated imaginary part of the spin susceptibility is shown in Fig.~4a for different hole concentrations. In the undoped case, one can observe two peaks in the imaginary part of the spin susceptibility: an intense peak corresponding to the single-magnon excitation, and a weaker feature at higher energy corresponding to higher-order processes. The single-magnon peak clearly disperses, although its energy is slightly above the value expected from linear spin-wave theory (dotted lines), due to the finite size of the cluster. In the inset of Fig.~4a, we present the energy-integrated magnetic intensity in the (H,K,0) reciprocal plane. As expected, the intensity for in the u
 ndoped situation is strongly peaked at ${\bf Q}_{AF}$, and rather uniformly distributed in the rest of the plane. As holes are added to the clusters, the magnetic spectral weight strongly decreases only around ${\bf Q}_{AF}$, but remains essentially constant everywhere else. This can also be seen in Fig.~4b, where the imaginary part of the energy-integrated spin susceptibility is plotted as a function of doping. The experimental and numerical data are thus in excellent agreement.

Armed with essentially complete knowledge of the spin fluctuation spectrum, and motivated by the agreement with the numerical data for the $t-J$ model, we now estimate the superconducting transition temperature $T_c$ generated by a Cooper pairing mechanism in which the experimentally detected spin fluctuations play the role of bosonic glue. To this end, we have self-consistently solved the Eliashberg equations using the vertex function of the $t-J$ model \cite{Prelovsek} and the experimental spin fluctuation spectrum of YBCO$_7$ (inset in Fig. 4c), without adjustable parameters. Details are given in the Supplementary Information. Figure 4c shows the outcome of the calculation. The resulting $T_c$ of 170 K is similar to the maximum $T_c$ observed in the cuprates and to another recent estimate based on the comparison of INS and photoemission data on underdoped YBCO$_{6+x}$ \cite{Dahm_NaturePhysics2009}. The agreement with the experimentally observed $T_c$ of YBCO$_7$ is satisfa
 ctory in view of the neglect of vertex corrections beyond Eliashberg theory and other simplifying assumptions in the calculation, and in view of a similar level of quantitative agreement reported for Eliashberg calculations of conventional low-$T_c$ superconductors.~\cite{Carbotte} In order to resolve the relative contribution of the strongly doping dependent region around ${\bf Q}_{AF}$ and the high-energy part of the spin susceptibility revealed by RIXS to the pairing strength, we have introduced momentum cutoffs around ${\bf Q}_{AF}$ in the calculation (Fig.~4c). Clearly, both low- and high-energy spin fluctuations contribute substantially to the pairing. In underdoped cuprates with lower gaps and stronger antiferromagnetic correlations, vertex corrections are expected to become essential~\cite{Schrieffer}, reducing the $T_c$ values calculated here. These corrections suppress coupling to magnons in the vicinity of ${\bf Q}_{AF}$, leaving the contribution of the higher-ene
 rgy excitations as a vital source of pairing.

\bibliographystyle{naturemag}

\bigskip

\noindent\textbf{Supplementary Information} accompanies this paper on www.nature.com/nphys

\smallskip

\noindent\textbf{Acknowledgements}
This work was performed at the ADRESS beamline using the SAXES instrument jointly built by Paul Scherrer Institut (Villigen, Switzerland), Politecnico di Milano (Italy) and \'{E}cole polytechnique f\'{e}d\'{e}rale de Lausanne (Switzerland). Part of this research project has been supported by the European Commission under the 7th Framework Programme: Research Infrastructures (Grant Agreement Number 226716), and the European project SOPRANO under Marie Curie actions (Grant No. PITNGA-2008-214040).

\smallskip

\noindent\textbf{Author Information} Correspondence and requests for materials should be addressed to
M.~L.~T. (\href{mailto:m.letacon@fkf.mpg.de}{m.letacon@fkf.mpg.de}) or B.~K.\ (\href{mailto:v.b.keimer@fkf.mpg.de}{b.keimer@fkf.mpg.de}).

%\clearpage
\pagebreak
\newpage
\begin{figure}
	\begin{center}
		\includegraphics[width=0.9\columnwidth]{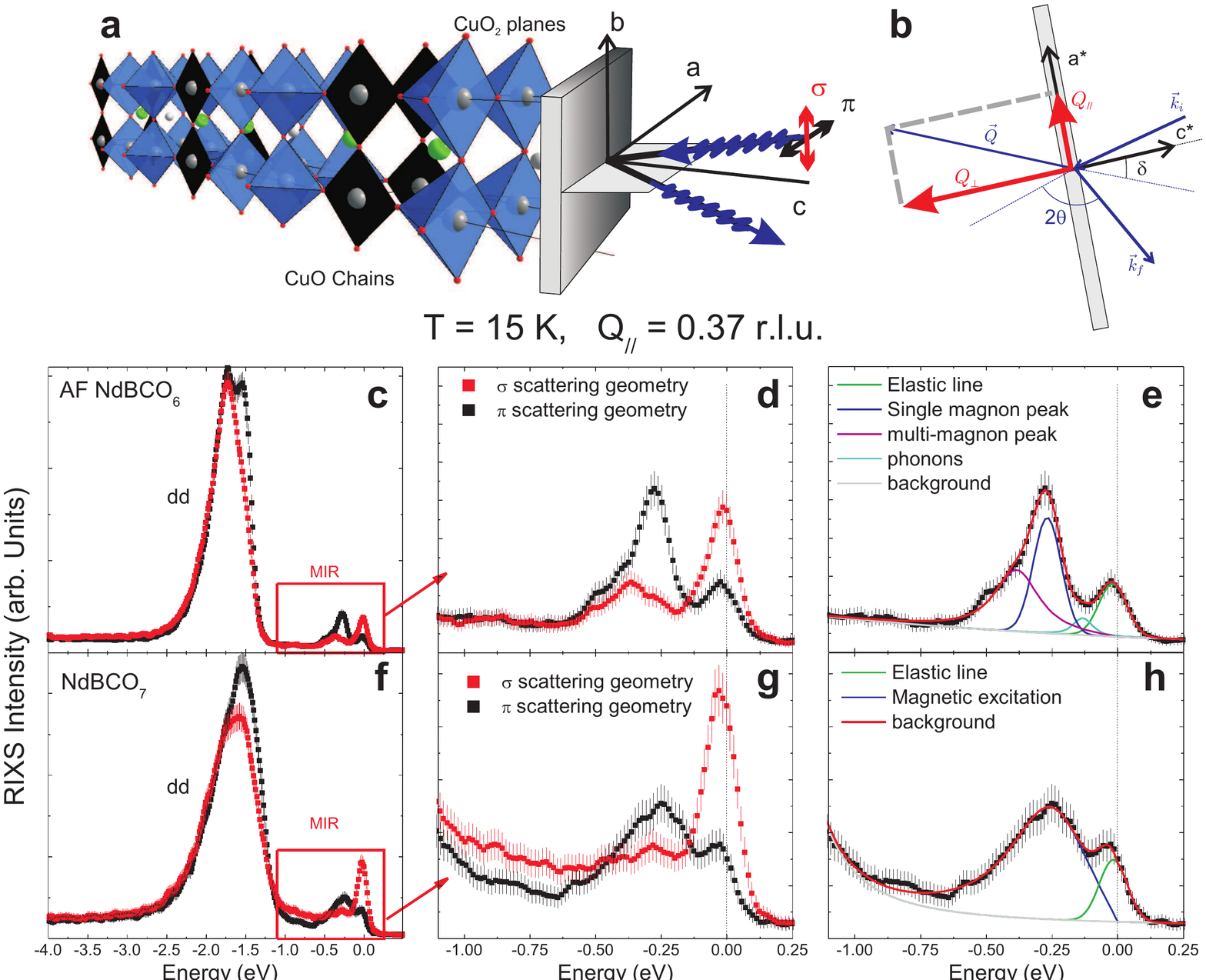}
	\end{center}
	\caption{a and b) Schematics of the scattering geometry. Typical RIXS spectra of undoped antiferromagnetic Nd$_{1.2}$Ba$_{1.8}$Cu$_3$O$_6$ (c, d, e) and superconducting underdoped Nd$_{1.2}$Ba$_{1.8}$Cu$_3$O$_{7}$ (f, g, h), obtained at T = 15 K for $q_{//}$=0.37 r.l.u., in both $\pi$ (black squares) and $\sigma$ (red squares) scattering geometries.}
	\label{fig:typical}
\end{figure}
\pagebreak
\newpage
\begin{figure}
	\begin{center}
		\includegraphics[width=\columnwidth]{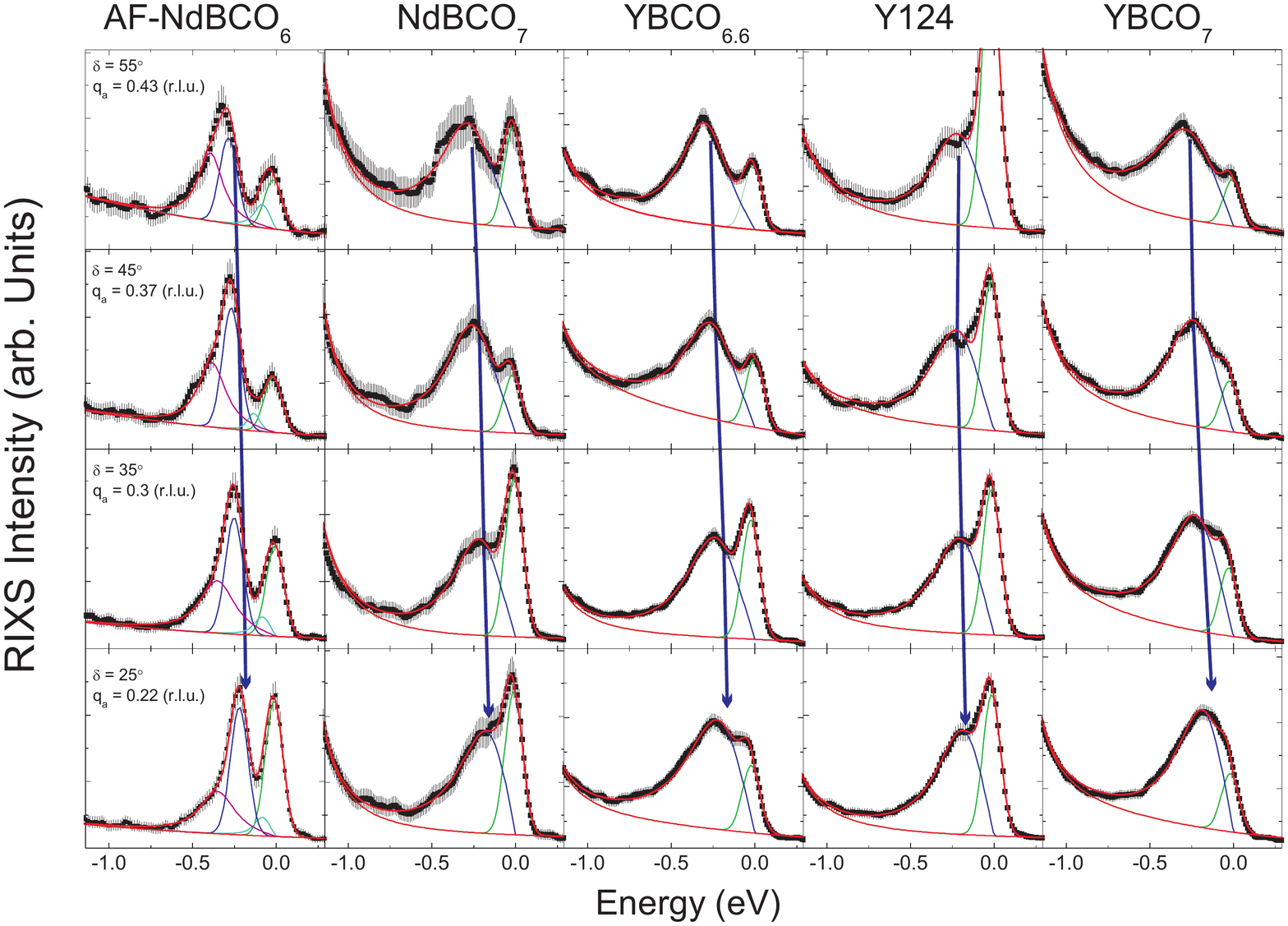}
	\end{center}
	\caption{RIXS response of undoped Nd$_{1.2}$Ba$_{1.8}$Cu$_3$O$_6$ (NdBCO$_6$), underdoped Nd$_{1.2}$Ba$_{1.8}$Cu$_3$O$_7$ (NdBCO$_7$), YBa$_2$Cu$_3$O$_{6.6}$ (YBCO$_{6.6}$), YBa$_2$Cu$_4$O$_8$, and slightly overdoped YBa$_2$Cu$_3$O$_{7}$ (YBCO$_7$) for various momentum transfers (see Fig.~1), at T = 15 K. Note that the spectra for NdBCO$_7$ were recorded for $\delta$=20$^\circ$, corresponding to $q_{//}$ = 0.18 r.l.u.}
	\label{fig:spectra}
\end{figure}
\pagebreak
\newpage
\begin{figure}
	\begin{center}
		\includegraphics[width=\columnwidth]{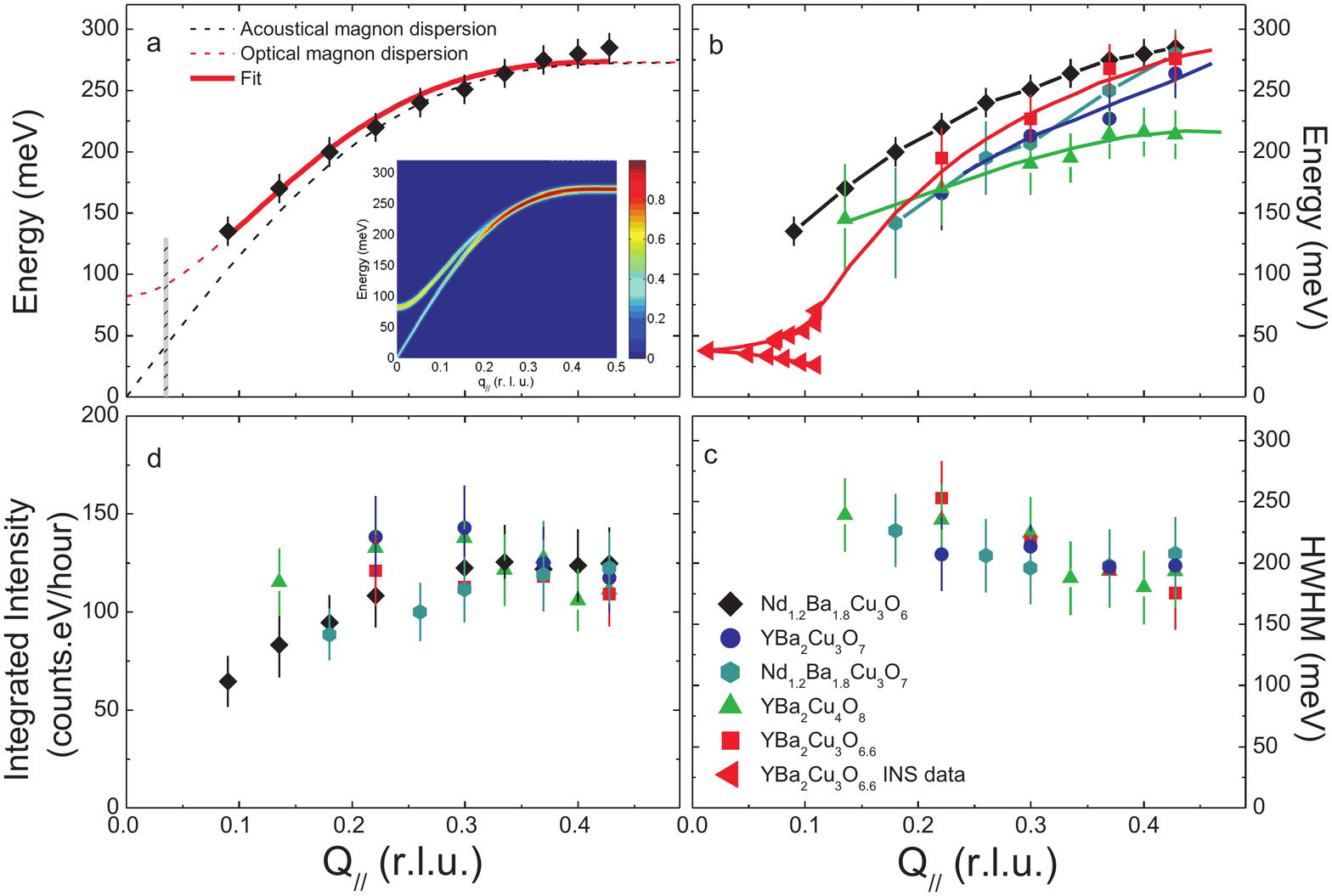}
	\end{center}
	\caption{a) Experimental magnon dispersion along 100 direction in AF Nd$_{1.2}$Ba$_{1.8}$Cu$_3$O$_6$ at T = 15 K, fitted using the spin-wave dispersion of a bilayer from Ref. \, \onlinecite{Tranquada_PRB1989} (thick red line).
The dashed lines are the acoustical (black) and optical (red) spin wave dispersions calculated using the fitting parameters. The grey area represents our energy/momentum resolution. Inset: relative intensity of the acoustical and optical magnon for our scattering geometry.
b) Experimental magnon dispersion along 100 direction in antiferromagnetic Nd$_{1.2}$Ba$_{1.8}$Cu$_3$O$_6$, underdoped Nd$_{1.2}$Ba$_{1.8}$Cu$_3$O$_7$, YBa$_2$Cu$_3$O$_{6.6}$, YBa$_2$Cu$_4$O$_8$ and YBa$_2$Cu$_3$O$_{7}$ at T = 15 K. Low-frequency INS data recorded along the 100 direction from $(\pi, \pi)$ for YBa$_2$Cu$_3$O$_{6.6}$ have been added \cite{Hinkov_natPhys2007}. Lines are guides to the eye. c) HWHM of magnetic excitations in Nd$_{1.2}$Ba$_{1.8}$Cu$_3$O$_7$, YBa$_2$Cu$_3$O$_{6.6}$, YBa$_2$Cu$_4$O$_8$ and YBa$_2$Cu$_3$O$_{7}$. d) Integrated inelastic intensities (see also Supplementary Information).}
	\label{fig:disp}
\end{figure}
\pagebreak
\newpage
\begin{figure}
	\begin{center}
		\includegraphics[width=0.8\columnwidth]{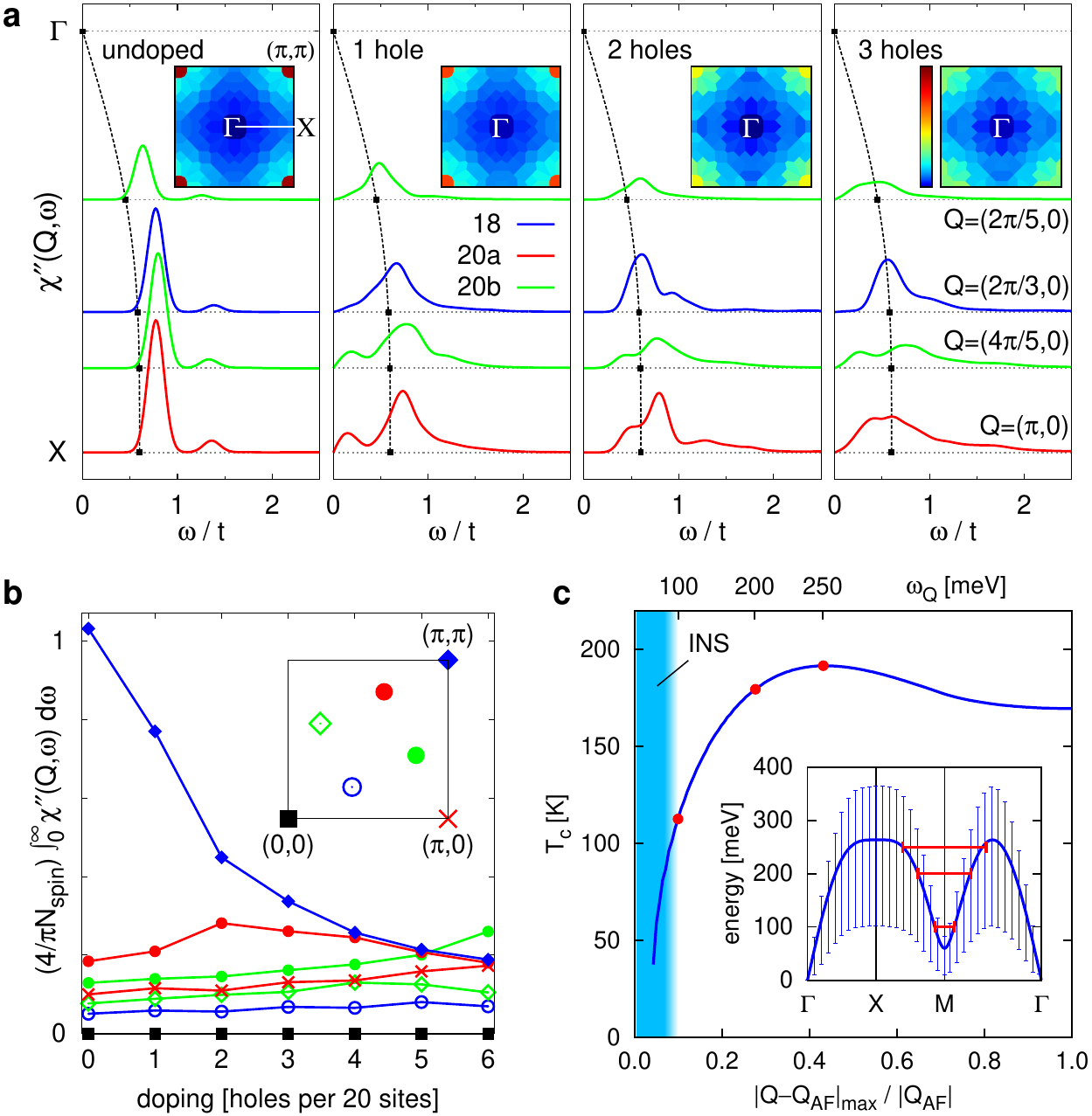}
	\end{center}
	\caption{a) Imaginary part of the spin susceptibility resulting from exact diagonalization of the $t-J$ model with $J/t=0.3$ on small clusters. Results
from 18-site and two 20-site clusters are combined to cover the $\Gamma-X$ line (see Supplementary Information). The spectra are broadened using a Gaussian with
HWHM$=0.1t$.  Dashed lines correspond to the linear spin wave dispersion with the same $J$ as used in numerics. Insets: Brillouin-zone-maps of energy-integrated $\chi''$ with a common color scale. b) Energy-integrated $\chi''$ of the 20-site cluster, normalized to the number of electrons on the cluster, as a function of doping. The seven accessible non-equivalent $\bf{q}$-vectors for this cluster are shown in the inset. c) Superconducting transition temperature resulting from the Eliashberg calculation for the experimentally determined spin excitation spectrum of YBCO$_7$ displayed in the inset. The red marks indicate the influence of successive momentum-space cutoffs limiting the maximum distance from ${\bf Q}_{AF}$ of the $\bf q$-vectors for which $\chi_q(\omega)$ is included in the
calculation.}
	\label{fig:tJ}
\end{figure}
\pagebreak
\newpage

\thispagestyle{empty}\clearpage
\section*{Supplementary Information for \\``Intense paramagnon excitations \\ in a large family of high-temperature superconductors''}
\subsection*{Sample Preparation and Characterization}
We used untwinned YBa$_2$Cu$_3$O$_{6.6}$ and YBa$_2$Cu$_3$O$_{7}$ single crystals of volume about $3 \times 3 \times 0.5$ mm$^3$, with superconducting transition temperatures $T_c$ of 61 K and 90 K ($\Delta T_c$ =2 K), respectively, determined for each crystal by SQUID magnetometry (see details in Methods section of Ref. \, \onlinecite{Hinkov_Nature04S}). The YBa$_2$Cu$_4$O$_8$ single crystals were obtained using KOH flux under ambient pressure in a box furnace. The source material was %used either from Y-124 ceramic powders or from
conventionally prepared polycrystalline YBa$_2$Cu$_3$O$_7$ mixed with CuO in a molar ratio of 1:1. The details are described in Refs. \, \onlinecite{Y124_a, Y124_b}. The crystal size was $0.8 \times 0.8 \times 0.2$ mm$^3$, and $T_c = 80$ K. Nd$_{1.2}$Ba$_{1.8}$CuO$_{6+x}$ films of thickness 100 nm were deposited on SrTiO$_3$ (100) single crystals by diode high-pressure oxygen sputtering.
The undoped Nd$_{1.2}$Ba$_{1.8}$CuO$_{6}$ film was obtained by annealing as-grown Nd$_{1.2}$Ba$_{1.8}$CuO$_{7}$ samples in Argon atmosphere (10 mbar) for 24 hours. Details of the film growth and characterization can be found in Ref. \, \onlinecite{Salluzzo}.

Table I lists the lattice parameters and transition temperatures of all samples.

\subsection*{Measurements}
The RIXS measurements were performed at the ADRESS beam line~\cite{Strocov_JSR2010} of the Swiss Light Source (Paul Scherrer Institute,
Switzerland) using the high-resolution SAXES spectrometer~\cite{Ghiringhelli_RSI2006}.
The resonant conditions were achieved by tuning the energy of the incident x-ray to the maximum of the Cu $L_3$ absorption peak, around 931 eV.
The total momentum transfer is 0.855 \AA$^{-1}$, which allows one to cover about 85$\%$ of the first Brillouin zone along the [100] direction (see Supplementary Figure 2).
Momentum transfers are given in units of the reciprocal lattice vectors $a^*$, $b^*$ and $c^*$ where $a^*$=2$\pi/a$, $b^*$=2$\pi/b$, and $c^*$=2$\pi/c$.
(See Table I for values of $a$, $b$, and $c$ for each sample).
The total energy resolution was about 130 meV, and the exact position of the elastic (zero energy loss) line was determined by measuring, for
each transferred momentum, a non-resonant spectrum of polycrystalline graphite.
All data where recorded at $T=15$ K, each spectrum being the result of 30 to 180 minutes total accumulation (sum of individual spectra of 5 min).

\subsection*{Data Analysis}
The data on the parent compound were analyzed using resolution-limited Gaussian lineshapes, as in Ref. \, \onlinecite{Braicovich_PRL2010S}.
Due to the broadening of the lineshape with doping, this analysis does not hold anymore for the other systems.
The RIXS intensity is proportional to the dynamical structure factor $S(\vec Q, \omega) = [1+n(\omega, T)] \chi^{\prime \prime} (\vec Q, \omega)$, where $[1+n(\omega, T)]=(1-e^{\frac{-\hbar \omega}{k_B T}})^{-1}$
is the Bose factor (step function at T = 15 K) and $\chi^{\prime \prime} (\vec Q, \omega)$ is the imaginary part of the spin susceptibility.
To take into account the finite lifetime of the excitations in the doped systems, and causality (that imposes $\chi^{\prime \prime} (\vec Q, \omega)$ to be an odd function of energy transfer $\omega$), we modeled $\chi^{\prime \prime} (\vec Q, \omega)$ using an antisymmetrized Lorentzian function, centered on $\omega=\omega_Q$, and of HWHM $\Gamma_Q$:
\begin{equation}
\label{eq:chi}
\chi^{\prime \prime} (\vec Q, \omega)= \bigg [ \frac{\Gamma_Q}{(\omega-\omega_Q)^2+(\Gamma_Q^2)} -\frac{\Gamma_Q}{(\omega+\omega_Q)^2+(\Gamma_Q^2)} \bigg  ],
\end{equation}
convoluted with the Gaussian resolution function (HWHM = 65 meV).
Note that in the case of doped systems, we are not able to separate the vibrational contribution to the inelastic response from the magnetic one. This may result in a slight, systematic overestimate of the intrinsic HWHM of the magnetic excitation plotted in Fig. 3c in the main text.
As in the undoped case, this contribution does not account for more than 10 \% of the total inelastic spectral weight (the inelastic intensities plotted in Fig. 3d all include this vibrational contribution).

In Supplementary Figure~1, the x-ray absorption spectra close to the Cu $L_3$ absorption edge (normal incidence, $\pi$ polarization) are shown for each of the investigated compounds. These measurements were taken on beamline ID08 at the European Synchrotron Radiation Facility (ESRF).
Since these absorption profiles are essentially identical between the different samples, the self-absorption corrections are not at all critical here.
The relative change of the intensities due to self-absorption obtained by doing calculations of the type given in Ref. \, \onlinecite{Eisebitt_PRB1993} are smaller than 3 \%. The high-energy $dd$ excitation peak (Fig. 1 b,c in the main text) provide an additional calibration standard.

\subsection*{Model calculations}

Supplementary Figure 3 shows the cluster shapes used in the numerical computation of the dynamical spin susceptibility of the $t-J$ model. The combination of different cluster shapes displayed in panel a provides access to the momentum points in panel b.

The Eliashberg calculations were performed with the vertex function of the $t-J$ model, $2J \gamma_q+(\epsilon_{k-q}+\epsilon_{k})/2$ where $\epsilon_k=-2t(\cos k_x+\cos k_y)-4t'\cos k_x\cos k_y$ represents the bare electronic dispersion and $\gamma_q=(\cos q_x + \cos q_y)/2$ \cite{PrelovsekS}. The values of the model parameters used in our calculation are $t=400$~meV, $t'=-t/3$, and $J=130$~meV.
As input we used the simple formula $\chi_q(\omega)=\eta_q/(\omega_q^2-\omega^2-i\Gamma_q\omega)$, which provides an excellent description of both the experimental data and the spin susceptibility resulting from the exact-diagonalization calculations. The parameters $\omega_q$ and $\Gamma_q$
denote the paramagnon dispersion and damping (full width at half maximum), respectively. We take $\omega_q=2J\sqrt{(1-\gamma_q)(1+\gamma_q+\omega_Q^2/8J^2)}$
with a spin gap $\omega_Q$ at ${\bf Q}_{AF}$. For underdoped YBCO$_{6+x}$, $\omega_Q$ can be directly extracted from INS data. For YBCO$_7$, where normal-state INS data are not available, the resonant mode below $T_c$ implies $\omega_Q \sim 60$ meV. This value is required to reproduce the experimentally observed resonant mode energy of 40 meV \cite{Fong96S} following method described in
Ref. \, \onlinecite{Prelovsek96S}. Further, $\eta_q=A(1-\gamma_q)$ with $A$ chosen such that $\chi''$ obeys the corresponding sum rule, as required by the RIXS and INS data.

Supplementary Figure 4 gives an overall picture of $T_c$ as a function of the damping
$\Gamma$ (chosen to be independent of $q$) and the spin gap at ${\bf Q}_{AF}$.
Apparently, non-damped spin waves with a suitably high energy at ${\bf Q}_{AF}$ are
optimal as the pairing mediators. $T_c$ decreases when the damping increases and/or when the
spin gap deviates from its optimum value, $\omega_Q \sim 100$~meV.

\pagebreak

\begin{table}
\begin{tabular}{|c|c|c|c|c|}
\hline
Sample  & a & b & c & Transition Temperature\\
\hline
\hline
Nd$_{1.2}$Ba$_{1.8}$Cu$_3$O$_6$ & 3.905 \AA & 3.905 \AA & 11.85 \AA & $T_N >$ 300 K\\
\hline
Nd$_{1.2}$Ba$_{1.8}$Cu$_3$O$_7$ & 3.885 \AA & 3.92 \AA & 11.7 \AA & $T_c$ = 65 K\\
\hline
YBa$_2$Cu$_3$O$_{6.6}$ & 3.82 \AA& 3.87 \AA & 11.7 \AA & $T_c$ = 61 K\\
\hline
YBa$_2$Cu$_4$O$_8$ & 3.84 \AA & 3.87 \AA & 27.25 \AA & $T_c$ = 80 K\\
\hline
YBa$_2$Cu$_3$O$_{7}$ & 3.817 \AA & 3.884\AA& 11.681\AA & $T_c$ = 90 K\\
\hline
\end{tabular}
\caption{\textbf{Supplementary Material} Lattice parameters and transition temperatures of the samples investigated here.}
\end{table}

\pagebreak

\begin{figure*}
	\begin{center}
		\includegraphics[width=\columnwidth]{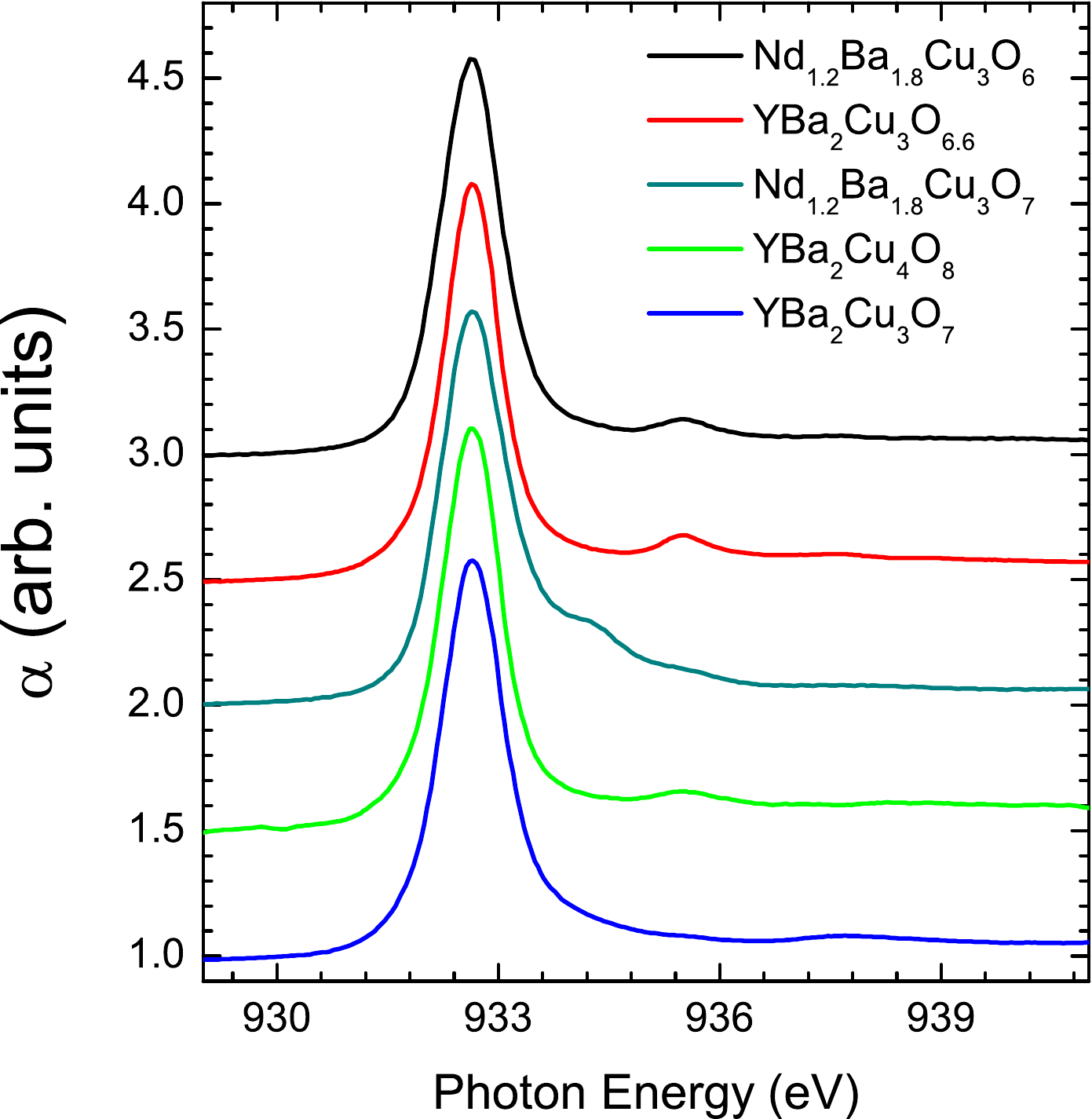}
	\end{center}
	\caption{\textbf{Supplementary Material} Cu $L_3$ edge X-ray absorption spectra of all the investigated compounds measured with total electron yield (normal incidence, $\pi$ polarization). Spectra have been vertically shifted by 0.5 for clarity.}
	\label{fig:XAS}
\end{figure*}

\pagebreak

\begin{figure}
	\begin{center}
		\includegraphics[width=\columnwidth]{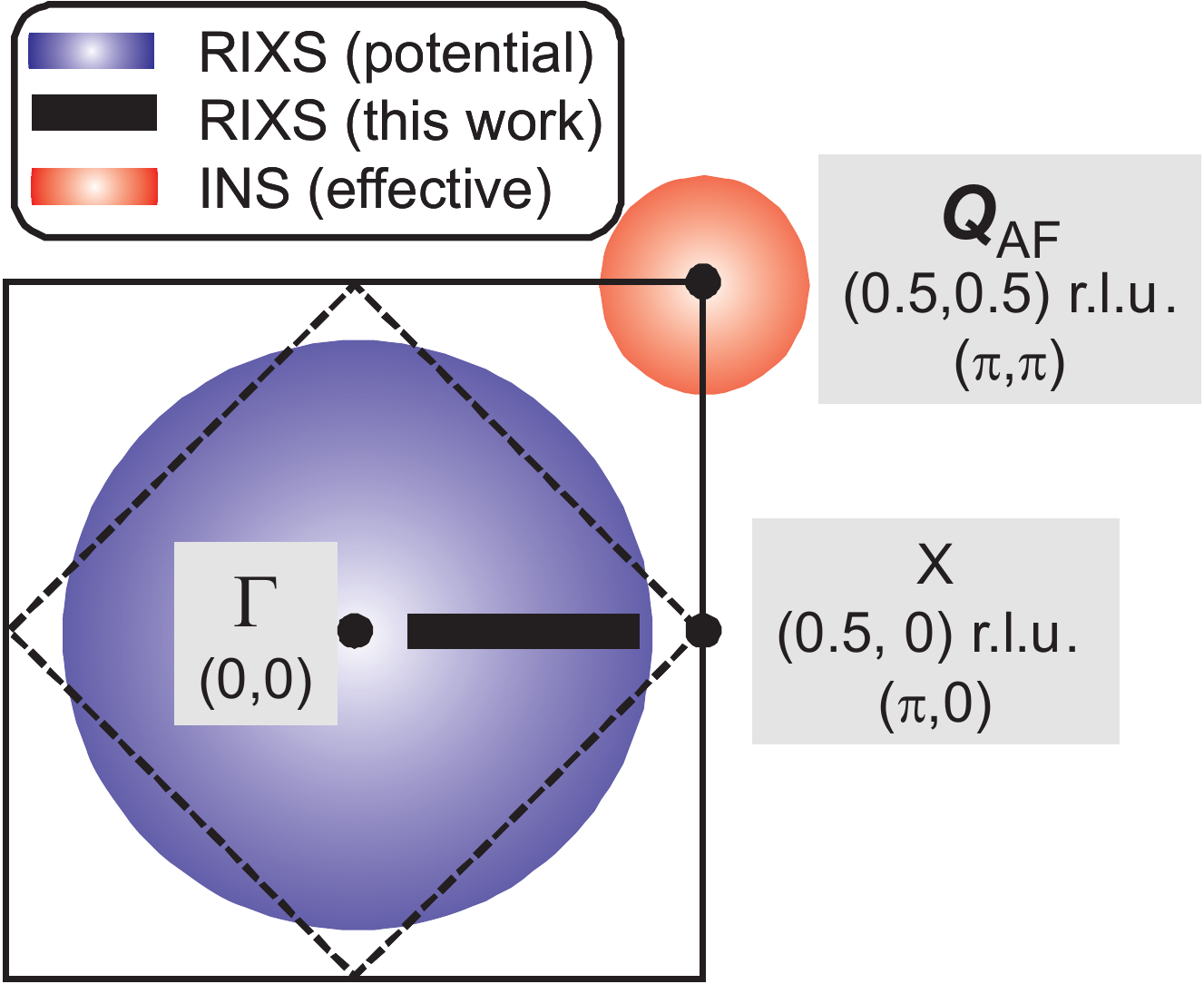}
	\end{center}
	\caption{\textbf{Supplementary Material} Sketch of the accessible reciprocal space with RIXS experiments at the Cu $L_3$ edge. The black line represents the region explored in this work. In red we have represented the typical region of the reciprocal space studied INS in doped YBCO.}
	\label{fig:XAS}
\end{figure}

\pagebreak

\begin{figure}
	\begin{center}
		\includegraphics[width=\columnwidth]{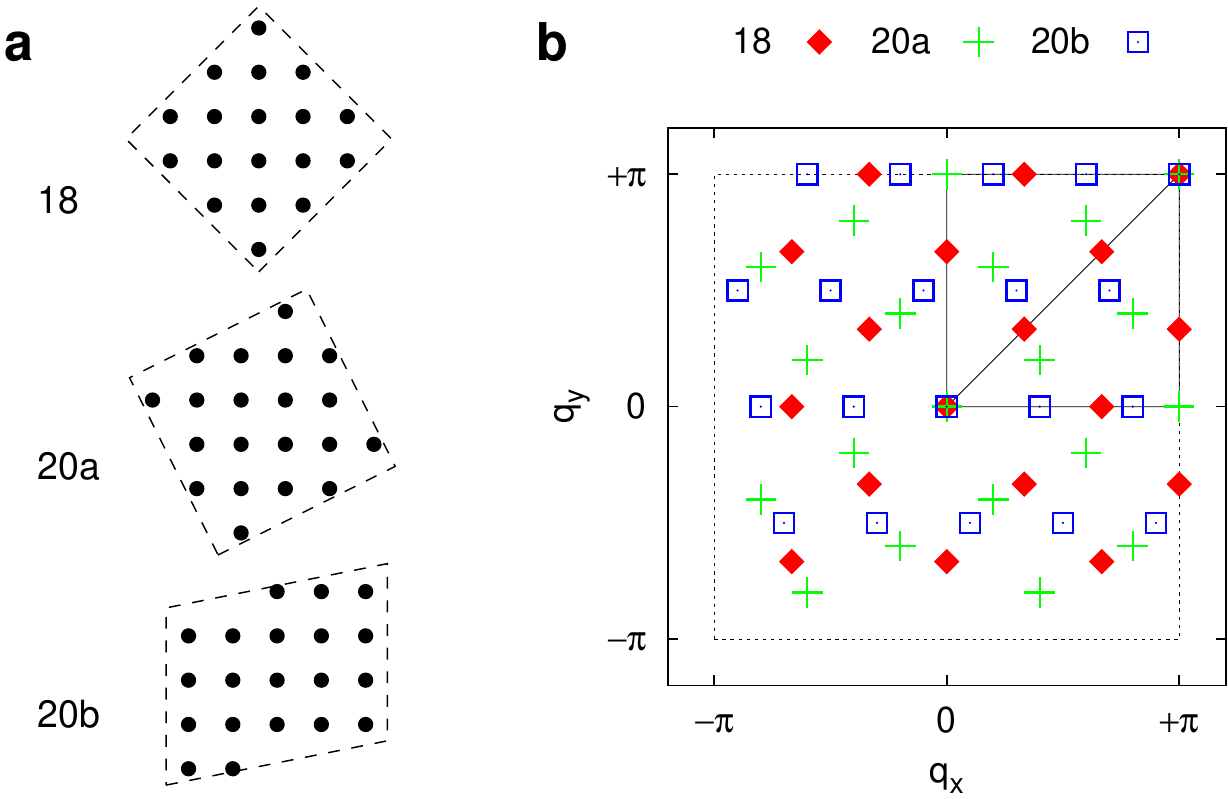}
	\end{center}
	\caption{\textbf{Supplementary Material} a) Cluster shapes used to compute the spin fluctuation spectra of Fig. 4 in the main text. b) Non-equivalent $\bf{q}$-vectors accessible to calculations on the different clusters.}
	\label{fig:clusters}
\end{figure}

\pagebreak

\begin{figure}
	\begin{center}
		\includegraphics[width=\columnwidth]{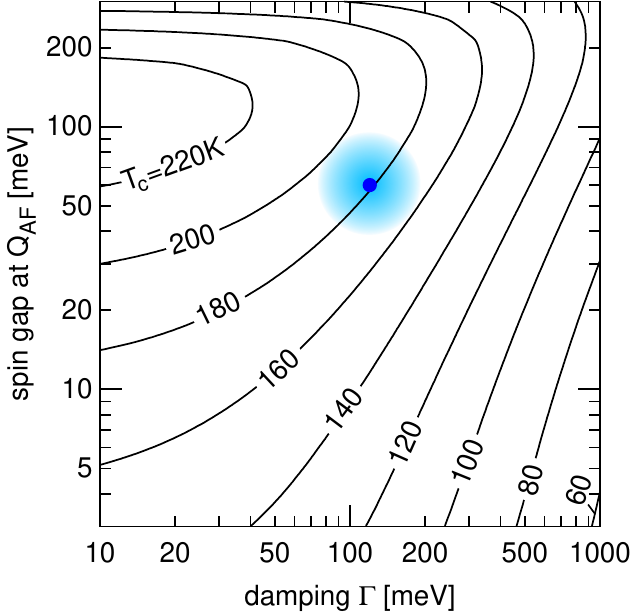}
	\end{center}
	\caption{\textbf{Supplementary Material} Contour plot of the superconducting transition temperature resulting from the Eliashberg calculation as a function of the spin gap at wave vector ${\bf Q}_{AF}$ and the damping parameter $\Gamma$. The values appropriate for YBCO$_7$ are marked in blue.}
	\label{fig:Tc}
\end{figure}

\end{document}